\documentclass[]{aa}       
\usepackage{graphicx}
\usepackage{amssymb}
\usepackage{amsmath}
\usepackage{longtable}
\usepackage{supertabular}
\usepackage{natbib} 

\begin{document}

\title{S5~0836+710: An FRII jet disrupted by the growth of a helical instability?}

\author{M. Perucho\inst{1} \and
             I. Mart\'{\i}-Vidal\inst{2} \and 
             A.P. Lobanov\inst{3}  \and
             P.E. Hardee\inst{4}
    }

\institute{ Dept. d'Astronomia i Astrof\'isica, Universitat de Val\`encia, 
C/ Dr. Moliner 50, 46100, Burjassot (Val\`encia), Spain 
\email{Manel.Perucho@uv.es}
\and
Onsala Space Observatory (Chalmers University of Technology),
Observatoriev\"agen 90, SE-43992 Onsala, Sweden
\and
Max-Planck-Institut f\"ur Radioastronomie (MPIfR),
Auf dem H\"ugel 69, D-53121 Bonn, Germany
\and
Department of Physics \& Astronomy, The University of Alabama, Tuscaloosa, AL 35487, USA
}

\authorrunning{Perucho et al.}

\titlerunning{An FRII jet disrupted by the growth of a helical instability?}

\offprints{M. Perucho, \email{manel.perucho@uv.es}}

\date{Received <date> / Accepted <date>}

\abstract
{The remarkable stability of extragalactic jets is surprising,
given the reasonable possibility of the growth of instabilities. In
addition, much work in the literature has invoked this possibility in
order to explain observed jet structures and obtain information from
these structures. For example, it was recently shown that the observed
helical structures in the jet in S5 0836+710 could be associated with
helical pressure waves generated by Kelvin-Helmholtz instability.}
{Our aim is to resolve the arc-second structure of the jet in the
quasar S5 0836+710 and confirm the lack of a hot-spot (reverse
jet-shock) found by present observing arrays, as this lack implies a
loss of jet collimation before interaction with the intergalactic medium.}
{In this work, we use an observation performed in 2008 using EVN
and MERLIN.  The combined data reduction has provided a complete image
of the object at arc-second scales.}
{The lack of a hot-spot in the arc-second radio structure is
taken as evidence that the jet losses its collimation between the VLBI
region and the region of interaction with the ambient medium.}
{This result, together with the previous identification of
the helical structures in the jet with helical pressure waves that grow
in amplitude with distance, allow us to conclude that the jet is
probably disrupted by the growth of Kelvin-Helmholtz instability. This
observational evidence confirms that the physical
parameters of jets can be extracted using the assumption that
instability is present in jets and can be the reason for many observed
structures. Interestingly, the observed jet is classified as a FRII
object in terms of its luminosity, but its large-scale morphology does
not correspond to this classification. The implications of this fact are
discussed.}

\keywords{galaxies: jets - hydrodynamics - instabilities - quasars: individual:
\object{S5~0836+710}}

\maketitle

\section{Introduction} \label{intro}

Jets in Active Galactic Nuclei (AGN) involve some of the most energetic
processes in the Universe. The way they 
propagate and interact with the ambient medium and, in particular, their
stability properties have been thoroughly 
studied \citep[see][for reviews]{ha06,ha11,pe11}. These jets are subject to the
growth of different instabilities and 
many of the structures observed like knots, bendings, helices, have been
interpreted as a result of this physical process. Although there could be other
ways to produce these structures, such as interactions with clumps of dense gas or
precession of the central engine, these in turn can also give rise to the
growth of the instabilities via coupling to any of the unstable modes
\citep[see, e.g.][for numerical studies of coupling to Kelvin-Helmholtz (KH) and current-driven (CD) instabilities 
in relativistic flows]
{pe+05,pe+06,mi07,mb09,mi09,mi10,pe+10,mi11}. Our interest in understanding how this process works in
jets lies in the possibility of obtaining the physical parameters of the flow
(velocity, gas density and sound speed), via modeling the observed structures
\citep[e.g.,][]{lz01,hr+05,he+11}. The Very Long Baseline Interferometry (VLBI)
technique has provided high-resolution observations that produce highly detailed
images of the structure of AGN jets, and that even resolves them transversally
\citep{lz01}. Aside from those studies in which it is assumed that the observed structures are generated by instabilities, 
no independent proof that this is indeed the case had been provided until \citet{pe+12}, where the authors showed that the 
ridge line of the jet in \object{S5 0836+710} behaves as expected if it is interpreted as a pressure wave generated by KH instability. 
They also showed, within this framework, that the amplitude of the wave grows with distance from the core, thus fulfilling the expected characteristics of an instability.  

The luminous quasar \object{S5 0836+710} at a redshift $z=2.16$ hosts a powerful
radio jet extending up to kiloparsec scales \citep{hu92}. At this redshift,
$1\,\rm{mas} \simeq 8.4\,\rm{pc}$ \citep[see MOJAVE database and][]{wr06}\footnote{https://www.physics.purdue.edu/astro/mojave/, which 
uses $\Lambda CDM$ Cosmology from WMAP 5 year results, and Ned Wrights Cosmology Calculator
http://www.astro.ucla.edu/\\$\sim$wright/CosmoCalc.html}. VLBI monitoring of the
source \citep{ot98} has yielded estimates of the bulk Lorentz factor
$\gamma_\mathrm{j}=12$ and the viewing angle $\alpha_\mathrm{j}=3^\circ$ of the
flow at milliarcsecond scales. The presence of an instability developing in the jet is suggested by the kink structures observed
on these scales with ground VLBI \citep{kr90}. \citet{lo98} observed
the source at 5~GHz with VSOP\footnote{VLBI Space Observatory Program, a Japanese-led 
space VLBI mission operated in 1997-2007 by the Institute of Space and Astronautical Science, Sagamihara, 
Japan http://www.vsop.isas.jaxa.jp/top.html} and also reported oscillations of the ridge-line.
Identifying these structures with KH modes, they were able to
derive an estimate of the basic parameters of the flow. High dynamic range VSOP
and VLBA (Very Long Baseline Array of National Radio Astronomy Observatory, USA)
observations of 0836+710 at $1.6\,\rm{GHz}$ indicated the presence of an
oscillation at a wavelength as long as $\sim100\,\rm{mas}$ \citep{lo06}, which
cannot be readily reconciled 
with the jet parameters given in \citet{lo98}. \citet{pl07,pl11} have shown that the
presence of a shear layer allows fitting all the observed oscillation
wavelengths within a single set of parameters, assuming that they are produced
by KH instability growing in a cylindrical outflow. In this picture, the
longest wavelength corresponds to a surface mode growing in the outer layers, whereas
the shorter wavelengths are identified with body modes developing in the inner
radii of the jet. In \citet{pe+12}, the authors have verified the relation between the helical structure and waves with
growing amplitude. 

In this paper we present a new observation of the jet in this source using EVN+MERLIN, following the experiment 
presented in \citep{hu92}. We confirm the observation of disrupted jet structure at arc-second scales and, using our 
present knowledge of the source, interpret this observation. 

\section{Observations}

The observations reported here were performed on 1 March 2008, and lasted 9.5 hours. 
We used the Multi-Element Radio-Linked Interferometer Network (MERLIN, United Kingdom)\footnote{MERLIN is a National Facility operated by the University of Manchester at Jodrell Bank Observatory on behalf of STFC} and, simultaneously, a subset of the European VLBI Network (EVN)\footnote{The European VLBI Network is a joint facility of European, Chinese, South African, and 
other radio astronomy institutes funded by their national research councils}.
On the one hand, there was a total of 9 participating antennas at the EVN: Effelsberg (100\,m diameter), Cambridge (32\,m), Lovell (76\,m), Onsala (25\,m), Medicina (32\,m), Noto (32\,m), Torun (32\,m), Urumqi (25\,m), and Shanghai (25\,m); on the other hand, there were 7 participating antennas at MERLIN (some of them simultaneously used as EVN stations): Defford (25\,m), Cambridge (32\,m), Knockin (25\,m), Darnhall (25\,m), MarkII (25$\times$38\,m), Lovell (76\,m), and Tabley (25\,m);  

The observations were performed at 1.6\,GHz. The recording rate per antenna at the EVN was 512 Mbps (full polarization, 8 frequency sub-bands of 8\,MHz each, 2-bit sampling). The overall synthesized bandwidth at the EVN was 64\,MHz, but the total bandwidth of MERLIN was limited to 16\,MHz. The EVN data were correlated at the Joint Institute for VLBI in Europe (JIVE, the Netherlands).

We used sources 4C39.25 and 3C345 as fringe finders, and B0859+681 as the amplitude/phase calibrator (3.9 degrees of separation). 
The initial amplitude and phase calibration was performed with the Astronomical Image Processing system (AIPS, NRAO\footnote{The National Radio Astronomy Observatory is a facility of the National Science Foundation operated under cooperative agreement by Associated Universities, Inc.}) using standard procedures. However, the phases of the different sub-bands in the EVN observations were aligned as described in Sect. 4.1 of \cite{mmVLBI}, due to non-constant sub-band phase offsets observed at Effelsberg and Urumqi.

\begin{figure*}[!t]
  \includegraphics[clip,angle=0,width=\textwidth]{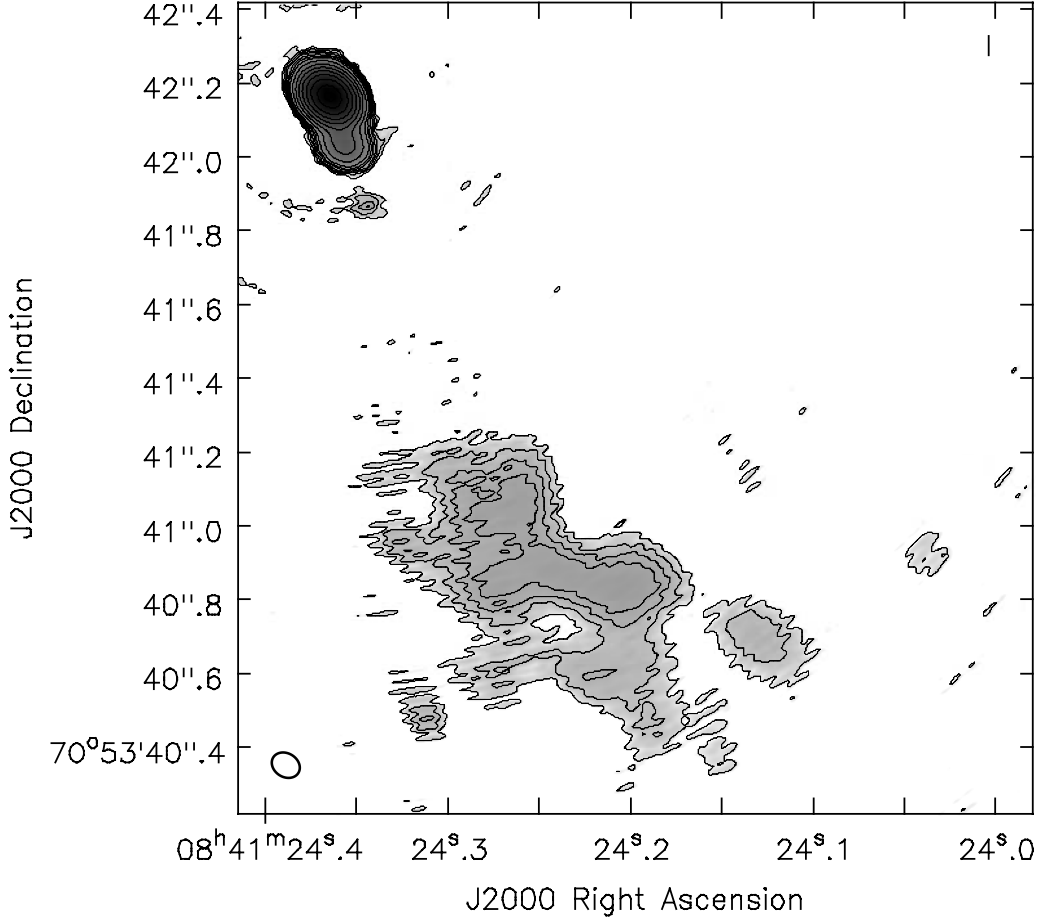}
    \caption{EVN+MERLIN image of \object{S5 0836+710} at 1.6~GHz. The contours are at 0.12, 0.15, 0.25, 0.33, 0.65, 1.2, 2.5, 5, 10, 20, 40, 80, and 99\% of the image peak intensity (2.48\,Jy/beam). The convolving beam is shown at the bottom-left corner. }
  \label{fig:map}
  \end{figure*}

Once the MERLIN and EVN observations were independently calibrated, the data were exported to the program CASA (Common Astronomy Software Applications, NRAO), version 3.3.0, for further calibration, reduction, and imaging. We used the \texttt{ms tool} of CASA to scale the amplitudes of the EVN visibilities to those of MERLIN at similar baseline lengths and imaged the combined dataset with the task \texttt{clean}, using {\em natural weighting} of the visibilities and a multi-scale deconvolution algorithm. We used beam deconvolution scales of 7\,mas (i.e., the size of the EVN synthesized beam), 170\,mas (i.e., the size of the MERLIN beam), 50\,mas, and 100\,mas (i.e., beams of intermediate size). We also used the \texttt{ms tool} to apply a phase gradient (i.e., a shift in the image plane) of about 1.5\,mas (south-west direction) to the EVN visibilities, to minimize the image residuals after the deconvolution (this is probably due to effects of source structure, which may introduce small offsets in the source intensity peaks at the very different resolutions achieved with MERLIN and the EVN). Our final image (with a convolving beam of 84$\times$68\,mas; position angle of 57 degrees)
 is shown in Fig. \ref{fig:map}.

\begin{figure}
  \includegraphics[clip,angle=0,width=\columnwidth]{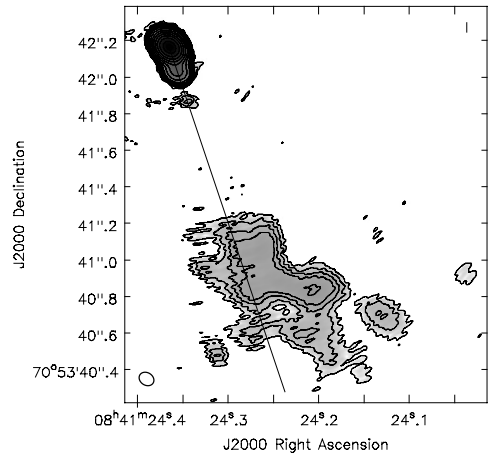}
    \caption{Same as Fig.~\ref{fig:map}, where a straight line following the VLBI jet direction has been drawn to help the eye.}
  \label{fig:line}
  \end{figure}

 The radio map presented in Fig.~\ref{fig:map} shows an irregular structure covering a region between 1 and 2 arc-seconds at the South-West from the radio core. A large portion of this region falls to the West of a straight line extension of the inner jet (see Fig.~\ref{fig:line}). 

\section{Discussion}\label{disc}

  The large-scale structure is observed at distances from the radio core larger than 1 arc-second, after the jet emission disappears at 0.2 arc-seconds. This implies an increase in the particle energy at the emission site. In typical FRII
jets this happens where the jet interacts with the ambient medium, typically the IGM, and a bright reverse jet-shock (hot-spot) is observed in many of these sources. At the hot-spot, the collimated jet flow is decelerated and its particles are deflected into the radio-lobes. The bow-like shape of the radio-lobes resembles the expected forward bow-shock in the ambient medium that is difficult to observe. 

  In the case of \object{S5 0836+710}, the images of the jet at the largest scales do not show any signs of enhanced emission that could be identified as the hot-spot or any clear lobe-like structure. On the contrary, the radio emission is irregular. The lack of a reverse jet-shock could imply the loss of jet collimation, as observed in FRI jets \citep[e.g.,][]{pm07}. After the loss of collimation, the jet is transformed into a subrelativistic or mildly relativistic broad flow \citep{pe+05} that continues propagating downstream and eventually interacts with the ambient medium. Possibly, the observed radio emission is then related to the acceleration of particles in the interaction region, which, depending on the flow velocity, could involve weak shocks.
  
 In the case of FRI jets, the deceleration and loss of collimation has been related to mass-loading from stellar winds \citep{lb02}, strong recollimation shocks \citep{pm07}, growth of nonlinear instabilities \citep{ro08} or inhomogeneities in the ambient medium \citep{mk08}. Observations seem to favor the first two possibilities in the region where FRI jets are efficiently decelerated, i.e., $\sim 1\,\rm{kpc}$. In the case of \object{S5 0836+710}, the VLBI jet is observed to keep collimation over more than 100 milli-arc-seconds, which at a viewing angle of 3$^\circ$ and $z=2.16$ implies $\sim 10-20\,\rm{kpc}$. At this distance, it seems difficult that mass-loading from stellar winds, even if accumulated along the path of the jet through the host galaxy, has anything to do with the loss of collimation. In addition, there are no signs of strong shocks along the VLBI jet. 
 
  Different authors have reported kinks in the jet at different scales (see the Introduction of this paper) and, in \citet{pe+12} the authors identify wave-like patterns in the observed helical structures. In addition, the amplitude of the helical motion increases with distance. 
These features are expected when an instability grows in a jet. Within this context, a small perturbation to the VLBI jet, namely a difference in pressure between two opposite sides of the jet, could couple to an unstable Kelvin-Helmholtz (KH, if the jet is matter dominated) or current driven (if magnetically dominated) mode that grows with distance to nonlinear amplitudes, which can lead to jet disruption and mixing with the ambient medium. There is a widely held belief that jets are dominated by particles beyond the most compact region, due to processes like mass-loading from stellar winds or clouds, or via the conversion of magnetic energy into kinetic energy of the particles in the process of jet acceleration \citep{ko09}. Thus, the KH scenario is favored.

 Linear stability analysis using expected parameters for the source \citep{pl07,pl11} shows that typical minimum growth lengths of disrupting, unstable modes such as the surface mode and the first body mode are $\lambda_i\sim 10^{2}\,R_j$. This distance provides the length after which the unstable mode amplitude is multiplied in an exponential factor. The 1.6~GHz images in \citet{pe+12} show that the VLBI jet is observed up to $\simeq 140\,{\rm mas}$ projected, which corresponds to $\simeq 22.5\,{\rm kpc}$ at a viewing angle of $3^\circ$. The observed growth in amplitude from the ridge line of the 1.6~GHz jet changes from $1-2\,{\rm mas}$ at $z<50\,{\rm mas}$ projected ($\simeq 8\,{\rm kpc}$ deprojected) to $4\,{\rm mas}$ at $z\simeq100-140\,{\rm mas}$ projected ($\simeq16-22.5\,{\rm kpc}$ deprojected). Taking into account that the amplitude of the instability evolves as $A(z)=A_0 exp(z/\lambda_i)$: 
 \begin{equation}
 A(z_2)=A(z_1)\frac{exp(z_2/\lambda_i)}{exp(z_1/\lambda_i)}.
 \end{equation}  
For $z_1=8\,{\rm kpc}$, and $z_2=24\,{\rm kpc}$ and using a typical jet radius of $R_j\simeq20\,{\rm mas}\simeq 170\,{\rm pc}$ at 1.6~GHz as obtained in \citet{pl07}, $A(z_2) = 4.7\,{\rm mas}$ if $A(z_1) = 2\,{\rm mas}$. Taking into account that the jet and ambient properties must change in such long distances and that this has a direct influence on the growth lengths of the unstable modes, the estimate is consistent with the observations. This consistency shows that the growth of the instability obtained from the linear analysis can explain in a self-consistent way the observed growth in the amplitude of the jet ridge-line at 1.6~GHz \citep{pe+12}.

 The growth of the instabilities is possible due to the continuous dissipation of kinetic energy from the jet flow so mild deceleration is expected to occur along the observed VLBI jet \citep[see, e.g.][]{pe+05,pe+10,he+11}. Deceleration and jet expansion would thus be the cause of the gap in emission between 0.2 and 1 arc-second, within our model.

The observational evidence obtained so far \citep[][and here]{pe+12} points clearly towards this physical process taking place in the jet of \object{S5 0836+710}. Several helical modes, including the more disruptive low order body modes, maybe the surface mode, and possibly some elliptical modes could be growing in the jet \citep{lo98,lo06,pe+12}. These wave modes generate helically twisted pressure waves, with a maximum that can be related to the peak of emission along the jet \citep[the ridge-line,][]{pe+12}. We conclude that this FRII jet is probably disrupted by the growth of a helical instability. This conclusion represents the first strong observational evidence of the development of instabilities in jet flows and validates many studies that have been made under the assumption that jets develop such instabilities \citep[e.g.,][]{har00,har03,hr+05,he+11}. 
 
   In addition, if this FRII jet is indeed disrupted by the growth of instabilities, the immediate question is whether there are other similar cases. If there are more cases, the morphological separation between FRI and FRII jets, depending on jet power, should be reconsidered accordingly. In particular, this could be a possible explanation for the different morphology of the jet and counter-jets in HYMORS \citep[hybrid morphology sources, e.g.,][]{gw00,gw06}, if one of the two jets is perturbed and the other is not. This requires asymmetric and irregular surrounding media (under the assumption that the jet and counter-jet in a source are symmetric), but moving clouds in the Broad Line and Narrow Line regions can interact with the jet laterally, inducing internal shocks or sound waves that can couple to unstable modes. The observations actually show helical structures in the FRI-like side of some sources \citep[see][]{gw06}. In the case studied here, it has been pointed out that the longest helical wavelength observed could be produced by long-term precession of the jet nozzle, with periods of the order of $T_{dr} \simeq 10^7$~yrs, which is a similar value to the one given by \cite{ha94}
for 3C~449. Future work should be focused on trying to verify that the growing instabilities are KH and to obtain as much information as possible from the linear perturbation theory. 

\begin{acknowledgements}
MP acknowledges
support by the Spanish ``Ministerio de Ciencia e Innovaci\'on''
(MICINN) grants AYA2010-21322-C03-01, AYA2010-21097-C03-01 and
CONSOLIDER2007-00050. PEH acknowledges support from NSF award AST-0908010 and
NASA award NNX08AG83G to the University of Alabama.

\end{acknowledgements}
\bibliographystyle{aa}

\end{document}